\documentclass[3p,times,twocolumn]{elsarticle}
\usepackage{ecrc}
\volume{00}
\firstpage{1}
\journalname{Nuclear Physics B Proceedings Supplement}
\runauth{}
\jid{nuphbp}
\jnltitlelogo{Nuclear Physics B Proceedings Supplement}
\usepackage{amssymb}
\usepackage[figuresright]{rotating}
\def\Journal#1#2#3#4{{#4}, {#1}, {#2}, #3} 

\begin{document}
\begin{frontmatter}
\dochead{}
\title{AMS-02 in space: physics results, overview, and challenges}
\author{Nicola Tomassetti, on behalf of the AMS collaboration}
\address{LPSC, Universit\'{e} Grenoble Alpes, IN2P3 / CNRS -- Grenoble, France}

\begin{abstract}
The Alpha Magnetic Spectrometer (AMS-02) is a state of the art particle detector measuring cosmic rays (CRs) 
on the International Space Station (ISS) since May 19th 2011. AMS-02 identifies CR leptons and nuclei in the 
energy range from hundreds MeV to few TeV per nucleon. 
Several sub-detector systems allow for redundant particle identification with unprecedented precision, a 
powerful lepton-hadron separation, and a high purity of the antimatter signal. The new AMS-02 leptonic data 
from $1$ to $500$ GeV are presented and discussed. These new data indicate that new sources of CR leptons 
need to be included to describe the observed spectra at high energies. Explanations of this anomaly may be 
found either in dark-matter particles annihilation or in the existence of nearby astrophysical sources 
of $e^{\pm}$. 
Future data at higher energies and forthcoming measurements on the antiproton spectrum and the boron-to-carbon 
ratio will be crucial in providing the discrimination among the different scenario.
\end{abstract}


\end{frontmatter}

\section{Introduction}     
\label{Sec::Introduction}  

\begin{figure}[!t]
  \centering
  \includegraphics[width=0.38\textwidth]{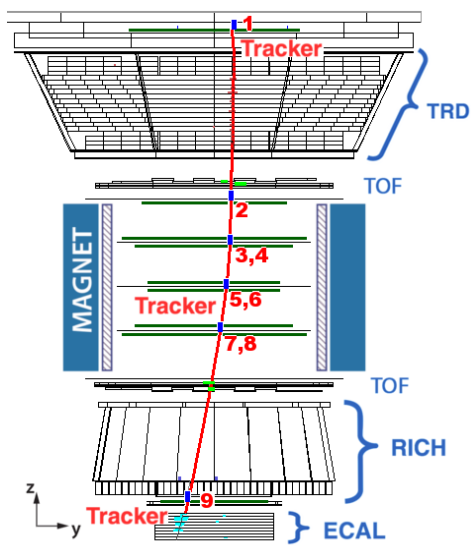}
  \caption{
    Schematic Y-Z view of the AMS-02 detector, illustrating the path of a typical CR particle crossing the various
    detector elements: the silicon Tracker, the TRD detector, the TOF scintillators, the RICH counter, and the ECAL.
    \label{Fig::ccAMSDetector}
}
 \end{figure}

AMS-02 is a general purpose high-energy particle detector, capable of measuring CR leptons and nuclei, 
from hydrogen up to iron, from hundreds MeV up to $\sim$\,1\,TeV of energy. 
AMS-02 was installed and activated on the ISS on 19 May 2011 to conduct a unique long duration 
mission ($\sim$\,20 years) of fundamental physics research in space. 
The layout of the AMS-02 detector is sketched in Fig.\,\ref{Fig::ccAMSDetector}. 
It consists of nine planes of precision Silicon Tracker, a Transition Radiation detector (TRD), four planes of Time of Flight counters (TOF), 
a permanent magnet, an array of AntiCoincidence Counters (ACC), a Ring imaging Cherenkov detector (RICH), 
and an Electromagnetic Calorimeter (ECAL).
The redundancy of the measurement ensures a correct particle identification and allows for detection of interactions inside the detector
\cite{Bib::TomassettiOliva2013, Bib::Saouter2013}
The TRD \cite{Bib::TRD} is built out of 5248 proportional chambers, filled with xenon and carbon-dioxide,
arranged in 20 layers with fiber--fleece radiator between each layer, allowing for discrimination between leptons and hadrons.
The TOF \cite{Bib::TOF} consists of four planes of scintillators, where two planes are located above the magnet case 
and two planes are placed below, allowing for velocity measurements, trigger, and charge measurements. 
The Tracker is build of nine planes of silicon micro-strip detectors \cite{Bib::Tracker}, from L1 to L9, 
distributed over the instrument. Planes from L2 to L8 are assembled with the permanent magnet, that has a magnetic field
strength of 0.15\,T. L1 is located on top of the TRD, and L9 is located between between RICH and ECAL. 
Each Tracker layer has a  spatial resolution of 10 $\mu$m.
The Tracker measures the rigidity $R$ up to $R\approx$2 TV ($R\equiv p/Z$, momentum to charge ratio) 
for charge one particles, and charge measurement up to $Z=28$.
The RICH \cite{Bib::RICH} is made of a radiator layer, a conical mirror, and a photomultiplier
plane for detecting the Cherenkov light. 
A NaF (aerogel) radiator with refractive index $n$=1.34 ($n$=1.04) is used in the central (outer) region to provide
high-precision measurement of the particle velocities and charge.
The ECAL \cite{Bib::ECAL} is a sampling calorimeter made out lead and scintillating fibers arranged in 16 layers.
The ECAL provides 17 radiation lengths of detecting medium, which allows for a precise reconstruction 
of the energy up to several TeV. A boosted decision tree (BDT) classifier is used to identify leptons
based on their 3D shower shapes.
Mounted on the ISS, AMS-02 is orbiting the Earth at an altitude of about 400 km, with inclination of 51.6$^{\circ}$.
The average trigger rate is 600 Hz with event size of 2 kByte. 
The minimal down-link bandwidth is 9 Mbit/s.
The detector is controlled from the AMS-02 Payload Operations Control Center (POCC) at CERN, Geneva. 
From the POCC, a continuous monitoring of the data flow and detector systems is performed.
This includes health and status of all sub-systems and the temperatures of the individual components.
Intensive time-dependent calibrations have been performed to all sub-detectors. 
No significant degradation of the sub-systems has been observed during 3 years of operation in the ISS.

\section{New results}  
\label{Sec::Results}   

AMS-02 collects about 1.5 billion CRs during each month of operation. 
The new results are based on the data collected during the initial 3 years of operations on the ISS, from May 2011 to May 2014. 
This constitutes $\sim$\,16\% of the expected AMS data sample.
In the measurement of leptons, the TOF is used to select $Z = 1$ relativistic particles traversing AMS-02 in the downward 
direction. The signals released in the TRD and ECAL detectors are used to discriminate the leptonic component from the hadronic background. 
A track reconstructed from L2 to L8 and matching the TRD and ECAL signals is used to select clean $Z = 1$ particles. 
For the discrimination of leptons from the dominant background of hadrons, three methods are used:
a TRD-likelihood based estimator, an ECAL-BDT estimator, and the $E/R$ ratio between ECAL energy and Tracker rigidity. 
This redundancy allowed to characterize the performance of each discrimination method using the data. 
Various analyses with different combinations of cuts or template fits were also tested as a cross-check.
Another important source of background comes from \emph{charge-confusion} of electron events. It may arise from to the finite 
rigidity resolution of the Tracker, or from the emission of secondary particles near the primary track.
The charge-confusion contribution can be estimated from data and calculated with Monte Carlo simulations. The relevant 
processes are well described by the simulations \cite{Bib::AMSPositronFraction2013}.
\begin{figure*}[htbp]
\centering
\includegraphics[width=0.43\textwidth]{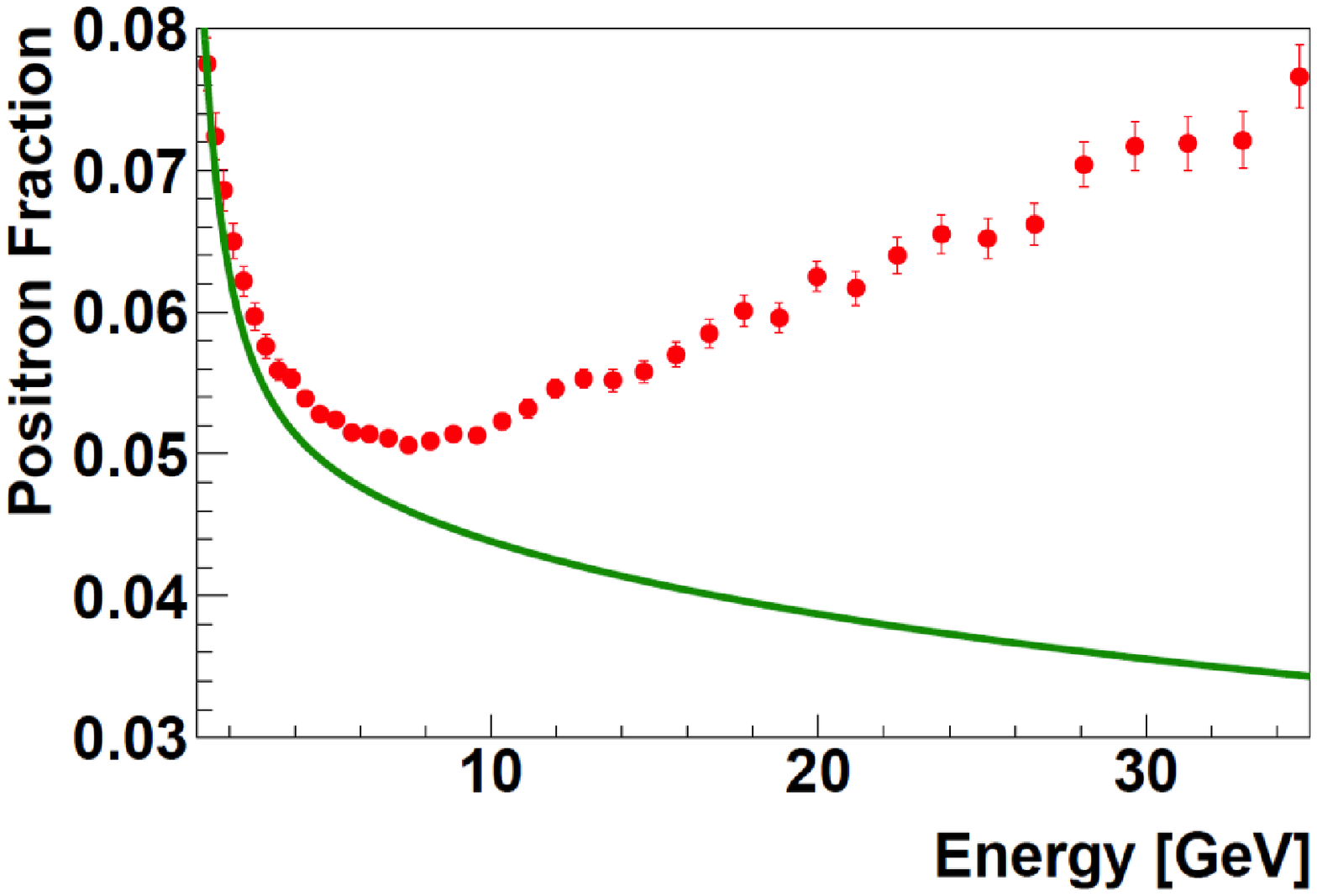}
\includegraphics[width=0.48\textwidth]{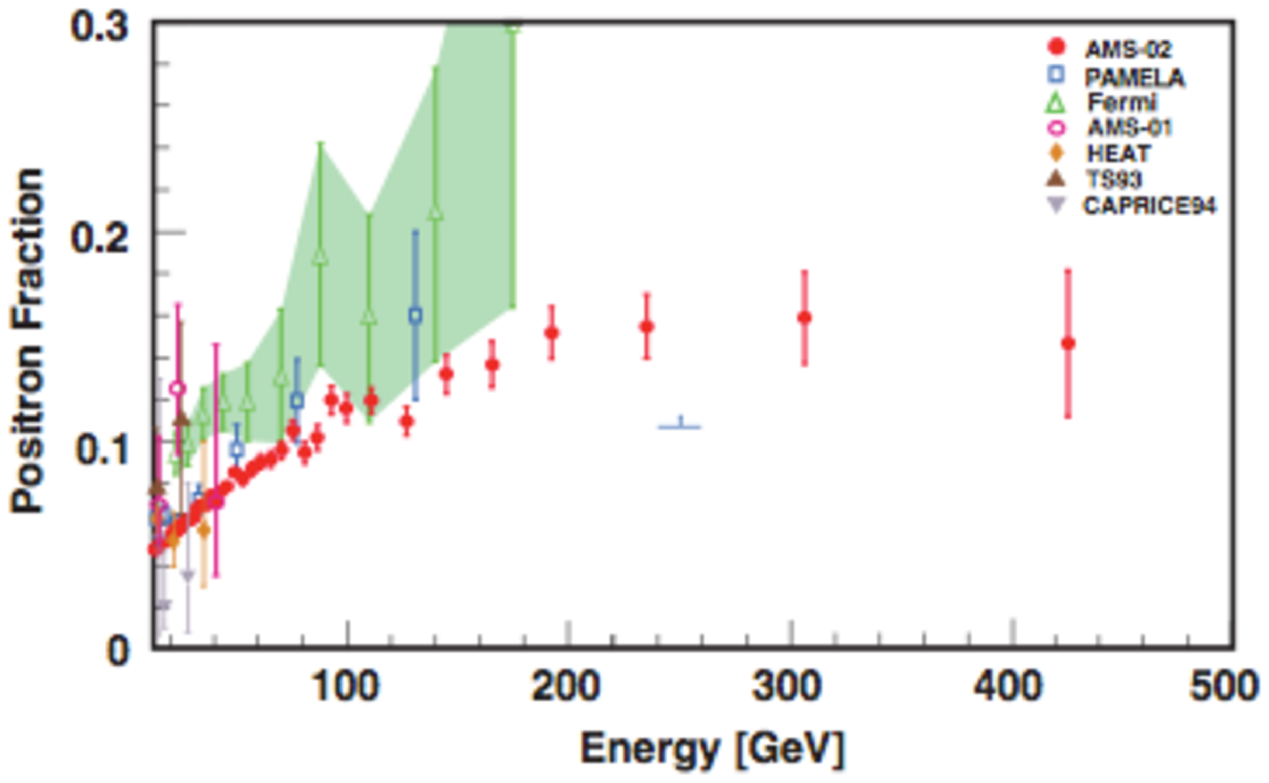}
\caption{AMS-02 positron fraction data from 1 to 35 GeV (left) and from 10 to 500 GeV (right) compared with previous measurements.}
\label{Fig::ccPositronFraction2014}
\end{figure*}
A new measurement of the positron fraction, $e^{+}/(e^{+} + e^{-})$, from 1 to 500 GeV of energy, 
has been recently presented \cite{Bib::AMSPositronFraction2014}. The results are shown in Fig.\ref{Fig::ccPositronFraction2014}.
The left figure shows the fraction up to 35 GeV of energy compared with the model prediction from
conventional calculations. In conventional models, CR electrons are emitted from supernova remnants (SNRs), 
while secondary $e^{\pm}$ arise from collisions of CR nuclei with the interstellar matter (ISM). 
From this model, the positron fraction must have a decreasing behavior with energy. Below few GeV the AMS 
data decrease with energy, as expected, although this energy region is 
affected by charge-dependent solar modulation. 
Above 10\,GeV, the data show a persistent rise up to 200 GeV, in clear contrast with the conventional picture, 
followed by an intriguing flattening at higher energies (right panel).
Additional information can be obtained from the single spectra of $e^{+}$ and $e^{-}$, that have been 
recently published \cite{Bib::AMSElectronAndPositron2014}.
The data show a slight spectral hardening of both $e^{-}$ and $e^{+}$ component above $\sim$\,30\,GeV, where
the $e^{+}$ spectrum experience a stronger hardening than the $e^{-}$ spectrum. Both spectra are
substantially smooth over the measured energy range. The data are also consistent with a 
new dedicated analysis of the total $(e^{-} + e^{+})$ spectrum up to $E=$\,1\,TeV of energy, 
where no charge-sign separation is performed \cite{Bib::AMSAllElectron2014}. 
All these data indicate the existence of a new source of $e^{+}$ and $e^{-}$ in 
addition to the conventional predictions based on secondary production. 
Such a source can be interpreted either in terms of dark matter (DM) particles annihilation
or in terms of nearby astrophysical sources of $e^{\pm}$.
The first class of interpretation requires DM particles with mass of the order of $\sim$\,1\,TeV, and may predict possible 
signals in the $\bar{p}/p$ ratio, depending on the DM--DM annihilation channels \cite{Bib::Boudad2014, Bib::Geng2014}. 
The present antiproton data show no clear evidence of such an excess within the uncertainties in the data and in the model predictions.
In the second class, a recent work demonstrated that the AMS-02 data may be described 
if nearby pulsar wind nebulae (PWNe) are accounted \cite{Bib::DiMauro2014}.
The PWN scenario gives no signatures in the antiproton channel.
It is also proposed that high-energy $e^{\pm}$ can be produced inside old-SNRs via interactions 
of CR protons with the background medium \cite{Bib::Blasi2009, Bib::MertschSarkar2014}. This mechanism 
This mechanism predicts remarkable signatures in the $\bar{p}/p$ ratio \cite{Bib::BlasiSerpico2009} as well as in the $B/C$ ratio 
\cite{Bib::MertschSarkar2009}, that may be detectable by AMS-02 \cite{Bib::TomassettiDonato2012}. 
Understandig the background due secondary production is therefore crucial for a multi-channel investigation of the CR data.
Present models are affected by large astrophysical uncertainties \citep{Bib::Tomassetti2012,Bib::Maurin2010}
that may be dramatically reduced with new data on CR (anti)protons and light nuclei. 

\section{Conclusions}    
\label{Sec::Conclusions} 

After one century from the discovery of CRs, thanks to AMS-02 we are eventually entering the era of precision 
astroparticle physics and we can search for new physics phenomena using CR data. 
High precision data from AMS-02 of the positron fraction and the electron and positron spectra
point consistently to the existence of a new source of high-energy leptons, 
that can be interpreted either by DM annihilation or by astrophysical objects such as PWNe or old-SNRs.
We emphasize the crucial role of the CR propagation physics in understanding the nature of such a new source.
First, for having a robust estimate of the level of \emph{astrophysical  background} from secondary production of $e^{\pm}$,
which is presently affected by large uncertainties. 
Second, for modeling the propagation effects on the spectral shape of both signal and background of $e^{\pm}$.
Third, for testing the different scenarios using CR nuclear data. For instance, high-energy measurements of 
CR antiprotons and boron-to-carbon ratio may provide a conclusive discrimination among the DM-scenarios  
(from which one may expect an excess in the $\bar{p}/p$ ratio), old-SNR scenarios (which predict signatures on the $B/C$ ratio) 
or PWN scenarios (from which no signatures are expected in the nuclear channels). 
Clearly, it is also crucial the behavior of the positron fraction in the high energy region. 
The behavior between $\sim$\,200 GeV and $\sim$\,1\,TeV will become more transparent with 
more statistics which will also allow improved treatment of the systematics.

\section{Acknowledgement}    
\label{Sec::Acknowledgement} 
This work is supported by acknowledged persons and institutions 
in \cite{Bib::AMSPositronFraction2013} and by the Labex grant \textsf{ENIGMASS}.


\begin{thebibliography}{00}
\bibitem{Bib::TomassettiOliva2013} Tomassetti, N., \& Oliva, A., 2013, 33$^{\rm th}$ ICRC, 896, Rio de Janeiro
\bibitem{Bib::Saouter2013} Saouter, P., et al., 2013, 33$^{\rm th}$ ICRC, 789, Rio de Janeiro
\bibitem{Bib::Tracker} Alpat, B., et al., \Journal{NIM}{A 613}{207}{2010}
\bibitem{Bib::TOF} Basili, A., et al., \Journal{NIM}{A 707}{99}{2013}
\bibitem{Bib::RICH} Aguilar, M., et al., \Journal{NIM}{A 614}{237}{2010} 
\bibitem{Bib::TRD} Kirn, T., et al., \Journal{NIM}{A 706}{43}{2013}
\bibitem{Bib::ECAL} Adloff, C., et al., \Journal{NIM}{A 714}{147}{2013} 
\bibitem{Bib::AMSPositronFraction2013} Aguilar, M., et al., \Journal{PRL}{110}{141102}{2013}
\bibitem{Bib::AMSPositronFraction2014} Accardo, L., et al., \Journal{PRL}{113}{121101}{2014}
\bibitem{Bib::AMSElectronAndPositron2014} Aguilar, M., et al.,\Journal{PRL}{113}{121102}{2014}
\bibitem{Bib::AMSAllElectron2014} Aguilar, M., et al., \Journal{PRL}{113}{221102}{2014}
\bibitem{Bib::Boudad2014} Boudad, M., 2014, arXiv:1410.3799
\bibitem{Bib::Geng2014} Geng, C. Q., et al., 2014, arXiv:1411.4450
\bibitem{Bib::DiMauro2014} Di Mauro, et al., \Journal{JCAP}{1404}{006}{2014}
\bibitem{Bib::Blasi2009} Blasi,~P., \Journal{PRL}{103}{051104}{2009}
\bibitem{Bib::MertschSarkar2014} Mertsch,~P., \&~Sarkar,~S., \Journal{PRD}{90}{061301}{2014}
\bibitem{Bib::BlasiSerpico2009} Blasi,~P., \& Serpico,~P.~D., \Journal{PRL}{103}{081013}{2009}
\bibitem{Bib::MertschSarkar2009} Mertsch,~P., \&~Sarkar,~S., \Journal{PRL}{103}{081104}{2009}
\bibitem{Bib::TomassettiDonato2012} Tomassetti,~N., \& Donato,~F., \Journal{A{\&}A}{544}{A16}{2012} 
\bibitem{Bib::Tomassetti2012} Tomassetti,~N., \Journal{Astrophys.Space Sci.}{342}{131-136}{2012}
\bibitem{Bib::Maurin2010} Maurin, D., et al., \Journal{A{\&}A}{516}{A67}{2010}
\end{thebibliography}
\end{document}